\documentclass[pdflatex,sn-mathphys-num]{sn-jnl}%
\usepackage{graphicx}%
\usepackage{multirow}%
\usepackage{amsmath,amssymb,amsfonts}%
\usepackage{amsthm}%
\usepackage{mathrsfs}%
\usepackage{graphicx}
\usepackage{subcaption}
\usepackage[title]{appendix}%
\usepackage{xcolor}%
\usepackage{textcomp}%
\usepackage{manyfoot}%
\usepackage{booktabs}%
\usepackage{algorithm}%
\usepackage{algorithmicx}%
\usepackage{algpseudocode}%
\usepackage{listings}%

\theoremstyle{thmstyleone}%

\theoremstyle{thmstyletwo}%

\theoremstyle{thmstylethree}%

\begin{document}

\title[Article Title]{Linguistic Predictability and Search Complexity:\\
How Linguistic Redundancy Constraints the Landscape of Classical and Quantum Search}

\author*[1]{\fnm{Alessio} \sur{Di Santo}}\email{alessio.disanto@graduate.univaq.it}
\equalcont{These authors contributed equally to this work.}

\author*[2]{\fnm{Gabriella} \sur{Lanziani}}\email{gabriella.lanziani@byteguardian.it}
\equalcont{These authors contributed equally to this work.}

\affil[1]{\orgdiv{Department of Information Engineering, Computer Science, and Mathematics}, \orgname{University of L’Aquila}, \orgaddress{\city{L'Aquila}, \postcode{67100}, \state{Abruzzo}, \country{Italy}}}

\affil[2]{\orgdiv{Independent Researcher}, \country{Italy}}

\abstract{
This study examines the quantitative relationship between linguistic regularities and computational search complexity through a hybrid classical--quantum framework applied to Renaissance Italian texts. Using four representative works from the fifteenth and sixteenth centuries—\emph{Il Principe} (Machiavelli), \emph{Il Cortegiano} (Castiglione), \emph{I Ricordi} (Guicciardini), and \emph{Orlando Furioso} (Ariosto)—we construct character-based $n$-gram models under both a historically grounded 25-letter orthography and the full modern Italian alphabet. These models provide corpus-derived probabilistic baselines for evaluating substitution-cipher search processes. Combining classical hill climbing and simulated annealing with Grover-style quantum-inspired estimates and a \textit{QUBO} annealing formulation, we quantify how the probability that a key produces a linguistically plausible decryption ($p_{\text{good}}$) relates to expected computational effort. Across cipher lengths from 200 to 1000 characters, empirical results confirm the predicted dependence of Grover oracle calls on $1/\sqrt{p_{\text{good}}}$ and show that longer texts yield sharper score distributions and smaller feasible key regions.

Overall, the findings establish a link between linguistic redundancy and search-space contraction, providing an empirical framework for comparing classical, quantum-inspired, and idealized quantum search dynamics under unified corpus-driven constraints.
}
\maketitle

\section{Introduction}\label{sec1}

Quantitative linguistics investigates the statistical organization of language and its implications for structure, variation, and processing \citep{Evidences,Balasubrahmanyan01121996}. Measures such as entropy, redundancy, and predictability describe how linguistic systems constrain the space of possible symbol sequences \citep{Altmann2005,Grzybek2007}. Since these measures quantify how sharply a corpus restricts linguistic variability, they offer a natural perspective for studying tasks that can be formulated as search problems.

A classical instance of such a search problem is the recovery of a monoalphabetic substitution cipher. A key corresponds to a permutation of the alphabet, and decrypting the text amounts to identifying the permutation that maximizes compatibility with a language model. The difficulty of this task depends on how strongly the language model concentrates probability mass around high-scoring permutations. Early work established the relevance of character and word statistics to cryptanalysis \citep{Shannon1951}, and later research has shown that these regularities persist across genres and historical periods \citep{Montemurro2014}.

In this study we explore this search setting through four representative works of Renaissance Italian—Machiavelli’s \emph{Il Principe} \citep{MachiavelliPrincipeWikisource}, Castiglione’s \emph{Il Cortegiano} \citep{CastiglioneCortegianoWikisource}, Guicciardini’s \emph{I Ricordi} \citep{GuicciardiniRicordiWikisource}, and Ariosto’s \emph{Orlando Furioso} \citep{AriostoFuriosoWikisource}. These texts span political philosophy, courtly dialogue, aphoristic reflection, and epic poetry. After orthographic normalization, we construct unigram, bigram, and trigram character models that provide probabilistic baselines for evaluating candidate decryptions.

Methodologically, we relate the probability that a randomly sampled key yields a linguistically plausible decryption ($p_{\text{good}}$) to the expected effort of several search paradigms:

\begin{itemize}
    \item classical hill climbing and simulated annealing,
    \item Grover-style quantum-inspired estimates based on $1/\sqrt{p_{\text{good}}}$,
    \item a \textit{QUBO} annealing formulation for permutation recovery.
\end{itemize}

This integrated approach allows us to examine how corpus-derived constraints manifest across classical and quantum-inspired search regimes.

Our main empirical findings are:

\begin{itemize}
    \item the distribution of $n$-gram scores becomes increasingly concentrated with longer cipher lengths, yielding a monotonic decline in $p_{\text{good}}$;
    \item the predicted Grover-style dependence on $1/\sqrt{p_{\text{good}}}$ holds across corpora and thresholds;
    \item classical heuristics and \textit{QUBO} annealing converge to similar high-scoring regions, indicating that linguistic structure strongly constrains the search landscape;
    \item despite substantial stylistic diversity, the four corpora exhibit closely aligned redundancy profiles at the character-trigram level.
\end{itemize}

The remainder of the paper reviews relevant work (Section~\ref{sec:background}), describes corpus preparation and modeling (Section~\ref{sec:methods}), presents the empirical results (Section~\ref{sec:results}), and discusses their implications for quantitative linguistics, computational search theory (Section~\ref{sec:discussion}) and final conclusions (Section~\ref{sec:conclusions}).

\section{Related work and theoretical background} \label{sec:background}
Research in quantitative linguistics has long examined how entropy, redundancy, and related measures capture structural properties of language \citep{Shannon1951,Montemurro2014,ZanetteMontemurro2005}. Such work shows that distributions of letters, words, and multiword units reflect grammatical and cognitive constraints that reduce the space of admissible sequences. Studies focusing on specific constructions, such as Japanese compound verbs \citep{Tamaoka2004}, demonstrate that these regularities can be highly sensitive to genre and register.

A second line of work investigates word-frequency distributions and lexical organization. Models based on Zipf-like behavior \citep{ZanetteMontemurro2005} and studies of frequency regimes \citep{FerrerCanchoSole2001} highlight how communicative pressures and combinatorial constraints shape emergent textual patterns. While informative, these approaches generally do not quantify how such regularities influence the complexity of search tasks.

A substantial cryptanalytic literature addresses automated decryption of monoalphabetic substitution ciphers. Early techniques relied on letter-frequency heuristics or genetic algorithms \citep{Gajek2011,article}, while later work refined scoring functions and search strategies through hill climbing, simulated annealing, or MCMC-based methods \citep{Bulatovic2019,Kambhatla2018}. These methods explore the language-model landscape induced by $n$-gram statistics, yet typically do not formalize the probability that a random key yields a plausible decryption.

Finally, quantum and quantum-inspired approaches have been explored for linguistic and combinatorial tasks. Grover’s search algorithm \citep{Grover1996} provides a quadratic speed-up for unstructured search, with expected iterations proportional to $1/\sqrt{p}$ for a marked fraction $p$ of solutions. Formal links between linguistic structure and quantum models have been investigated in compositional semantics \citep{Coecke2010} and quantum machine learning \citep{Eisinger2025}, but empirical tests of Grover-style scaling on corpus-derived distributions remain limited.

Table~\ref{tab:related_work} summarizes these strands and situates the present study within this landscape. Our contribution is to integrate corpus-driven estimates of $p_{\text{good}}$ with classical, quantum-inspired, and idealized quantum cost models, thereby linking redundancy in Renaissance Italian to measurable search-space contraction.

\begin{table}[t]
    \centering
    \caption{Overview of related work and how the present study extends existing research.}
    \label{tab:related_work}
    \begin{tabular}{p{2cm}p{2cm}p{3.8cm}p{4cm}}
        \toprule
        \centering \textbf{Domain} & \centering \textbf{Representative works} & \centering \textbf{Main focus} &\textbf{Extension in this study} \\
        \midrule
    
        Quantitative linguistics: entropy and redundancy &
        Shannon \citep{Shannon1951}; Montemurro \& Zanette \citep{Montemurro2014}; Tamaoka et al.\ \citep{Tamaoka2004} &
        Estimation of entropy, redundancy, and ordering regularities for natural languages. &
        Uses comparable $n$-gram statistics on Renaissance Italian, linking them to a concrete search task (cipher recovery) via $p_{\text{good}}$ and its dependence on text length. \\
        
        \midrule
        Statistical models of text structure &
        Zanette \& Montemurro \citep{ZanetteMontemurro2005}; Ferrer-i-Cancho \& Solé \citep{FerrerCanchoSole2001} &
        Zipf-like word-frequency behaviour and lexical organisation. &
        Demonstrates how these distributional constraints compress the effective key space and influence search complexity. \\
        
        \midrule
        Automated cryptanalysis of substitution ciphers &
        Gajek \citep{Gajek2011}; Bulatović et al.\ \citep{Bulatovic2019}; Kambhatla et al.\ \citep{Kambhatla2018}; Jain et al.\ \citep{article} &
        Heuristic, evolutionary, and neural optimisation techniques for accurate decryption. &
        Quantifies the fraction of linguistically plausible keys ($p_{\text{good}}$) and relates it to expected classical trials and Grover-style oracle costs. \\
        
        \midrule
        Quantum and quantum-inspired approaches to language processing &
        Grover \citep{Grover1996}; Coecke et al.\ \citep{Coecke2010}; Eisinger et al.\ \citep{Eisinger2025} &
        Application of quantum search, compositional semantics, and quantum machine learning to linguistic tasks. &
        Performs an empirical Grover-style analysis on historical corpora, testing oracle-call scaling and comparing quantum-inspired and classical optimisation under a unified scoring function. \\
        
        \bottomrule
    \end{tabular}
\end{table}

\section{Data and methods}
\label{sec:methods}

\subsection{Corpora}
We examine four Renaissance Italian works representing distinct genres. For each text, we train a separate language model to capture author-specific patterns. We also construct a merged corpus by concatenating all works, providing an aggregated view of Renaissance Italian.

\newpage

\subsection{Orthographic normalization}
Two parallel versions of each text were prepared:

\begin{itemize}
    \item a historically grounded 25-letter alphabet, standardizing \textsc{i/j}, \textsc{u/v}, and excluding letters such as \textsc{k}, \textsc{w}, and \textsc{y};
    \item the full modern Italian alphabet.
\end{itemize}

\begin{figure}[t]
    \centering
    \includegraphics[width=0.4\linewidth]{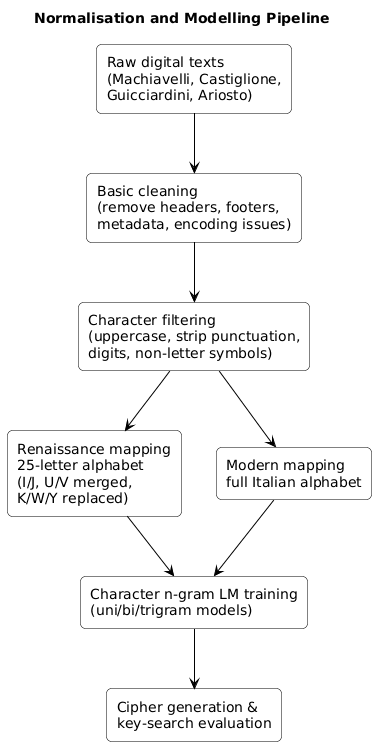}
    \caption{Normalization and modeling pipeline}
    \label{fig:normalisation-pipeline}
\end{figure}

Normalization follows conventions for early modern Italian corpora \citep{ITALIcamena,ManualeSNS,FontiPalio}. All diacritics, punctuation, and digits were removed, and text was converted to uppercase. This dual representation allows us to assess the robustness of statistical patterns across orthographic conventions (Fig.~\ref{fig:normalisation-pipeline}).

\subsection{Language model construction}
Character-based unigram, bigram, and trigram models were built for each text. For a sequence $c_{1:T}$, the $n$-gram model is

\begin{equation}
P(c_{1:T}) \approx \prod_{t=1}^{T} P(c_t \mid c_{t-n+1:t-1}),
\end{equation}

with $n=3$ used for most experiments. Trigram models offer a strong balance between expressiveness and statistical robustness for medium-sized historical texts, capturing short-range orthographic and phonotactic dependencies while avoiding the sparsity issues characteristic of higher-order models \citep{JurafskyMartin2025Ch3,Megyesi2023,SiivolaPellom2005}. Add-$\alpha$ smoothing ($\alpha=0.001$) was applied to avoid zero probabilities \citep{ChenGoodman1999}.

\subsection{Cipher generation}
Plaintext segments of lengths
\begin{equation}
L \in \{200,400,600,800,1000\}
\end{equation}
were sampled uniformly from each corpus and encrypted under uniformly random monoalphabetic substitutions. No word boundaries were preserved.

\subsection{Scoring function}
Each candidate key $\pi$ was evaluated by the log-likelihood of the decrypted text:

\begin{equation}
    S(\pi) = \log_{10} P_{\text{LM}}(\pi(c_{1:L}))
\end{equation}

This score landscape serves as the objective function for all optimization procedures.

\subsection{Estimating the fraction of good keys}
\begin{figure}
    \centering
    \includegraphics[width=0.85\linewidth]{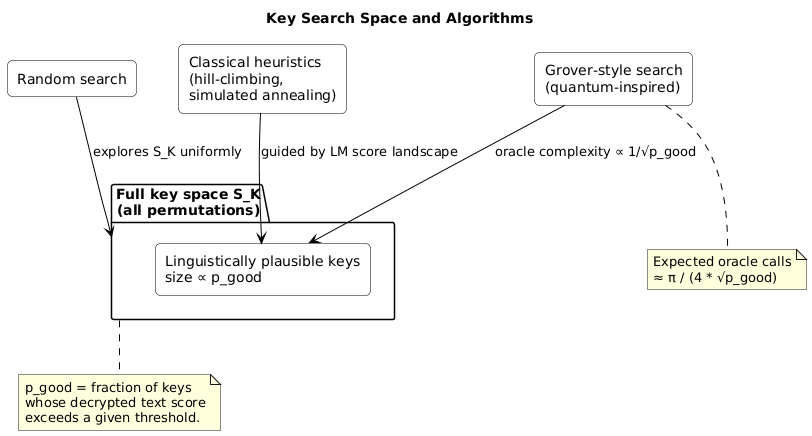}
    \caption{Schematic representation of the key search space.}
    \label{fig:search-space}
\end{figure}

For each corpus, length, and threshold $\tau \in \{0.90,0.92,0.95,0.98,0.99,0.995\}$, we estimated

\begin{equation}
p_{\text{good}} = \Pr(S(\pi) \geq \tau S_{\max}),
\end{equation}
where $S_{\max}$ is the highest score observed across all methods $\tau \in (0,1)$ defines the score level above which a candidate key is considered a \emph{marked} (or \emph{good}) solution in the sense of Grover’s search \citep{Grover1996}. Estimates were obtained via \textit{Monte Carlo sampling} over 10\,000 random permutations. Fig.~\ref{fig:search-space} illustrates the conceptual relation between the full permutation space and the subset of linguistically admissible keys.

\subsection{Classical optimization baselines}
We used two classical heuristic search methods:

\begin{itemize}
    \item \textbf{Hill climbing}: greedy local exploration via letter swaps with multiple restarts.
    \item \textbf{Simulated annealing}: stochastic exploration with geometric temperature decay
    \begin{equation}
        T_{t+1} = 0.99\, T_t.
    \end{equation}
\end{itemize}

The temperature $T$ controls the probability of accepting uphill moves, enabling the search to escape local maxima \citep{Kirkpatrick1983,AartsKorst1989}.

\begin{figure}[b]
    \centering
    
    \begin{subfigure}[t]{0.32\textwidth}
        \centering
        \includegraphics[width=\linewidth]{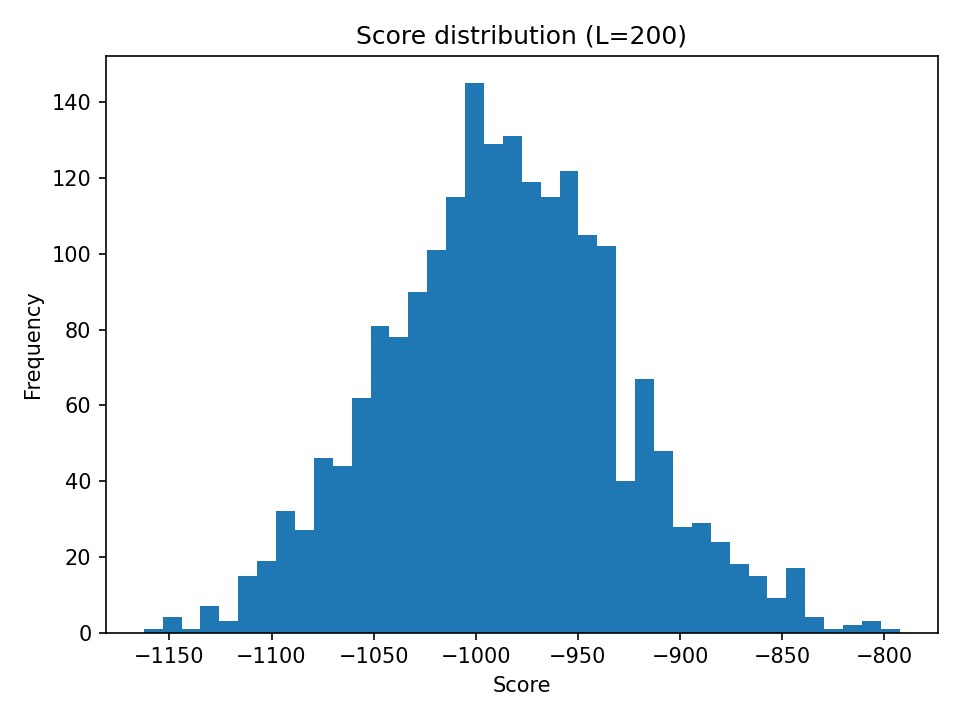}
        \caption{L = 200}
    \end{subfigure}
    \hfill
    \begin{subfigure}[t]{0.32\textwidth}
        \centering
        \includegraphics[width=\linewidth]{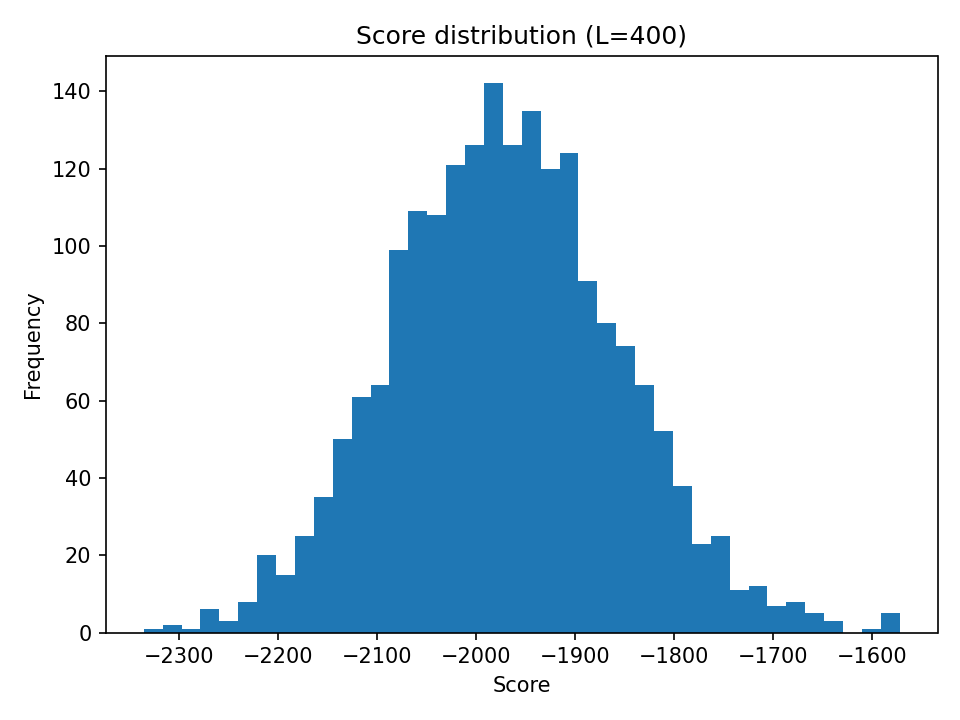}
        \caption{L = 400}
    \end{subfigure}
    \hfill
    \begin{subfigure}[t]{0.32\textwidth}
        \centering
        \includegraphics[width=\linewidth]{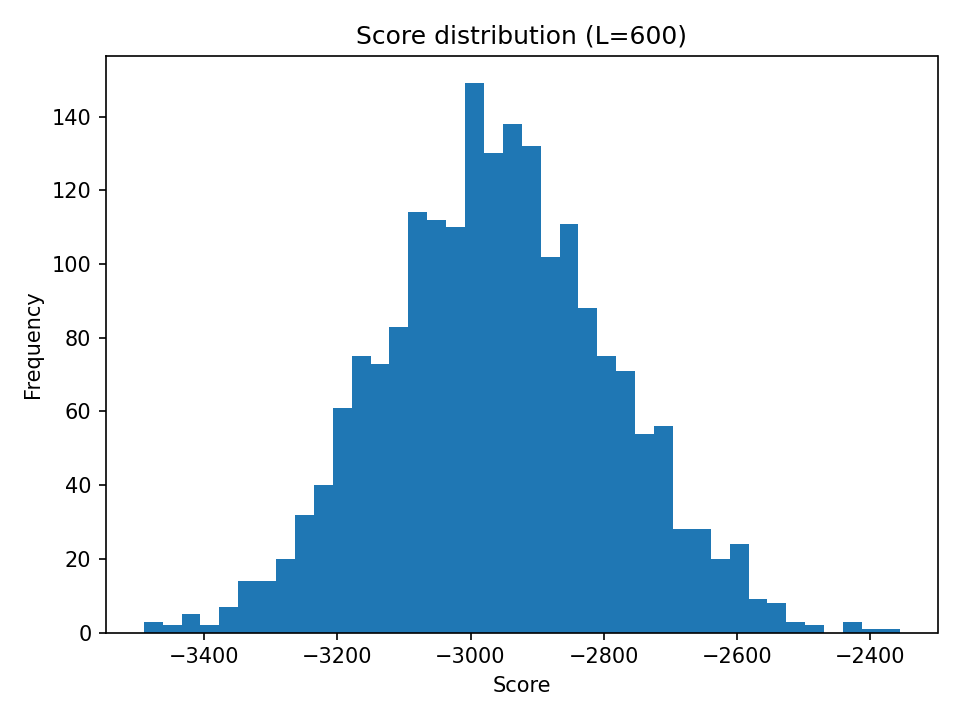}
        \caption{L = 600}
    \end{subfigure}
    
    \vspace{0.7em}
    
    \begin{subfigure}[t]{0.32\textwidth}
        \centering
        \includegraphics[width=\linewidth]{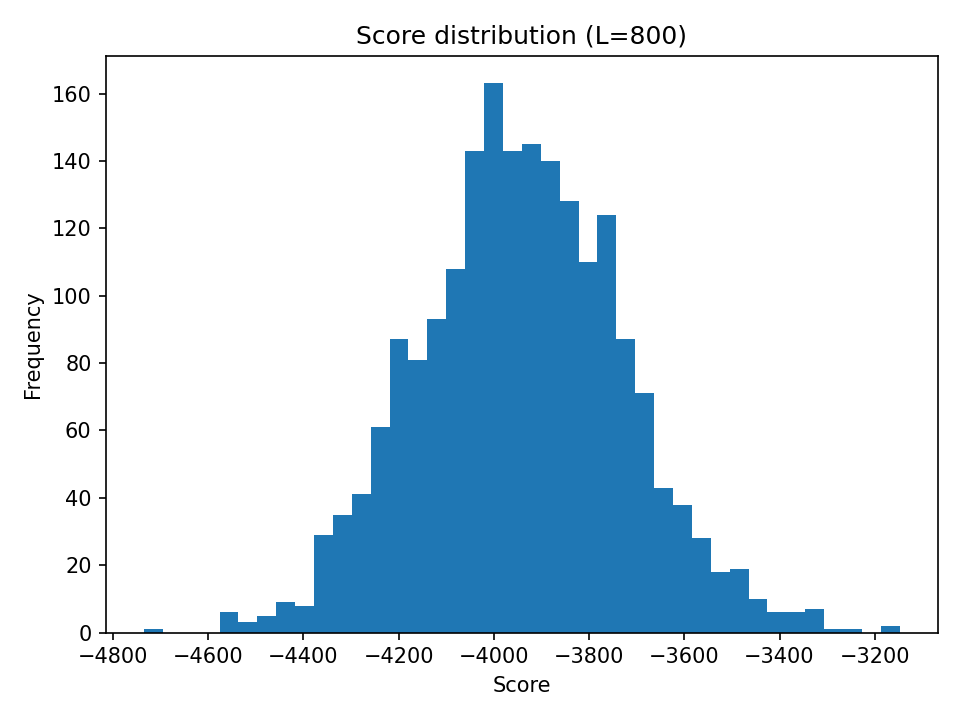}
        \caption{L = 800}
    \end{subfigure}
    \hfill
    \begin{subfigure}[t]{0.32\textwidth}
        \centering
        \includegraphics[width=\linewidth]{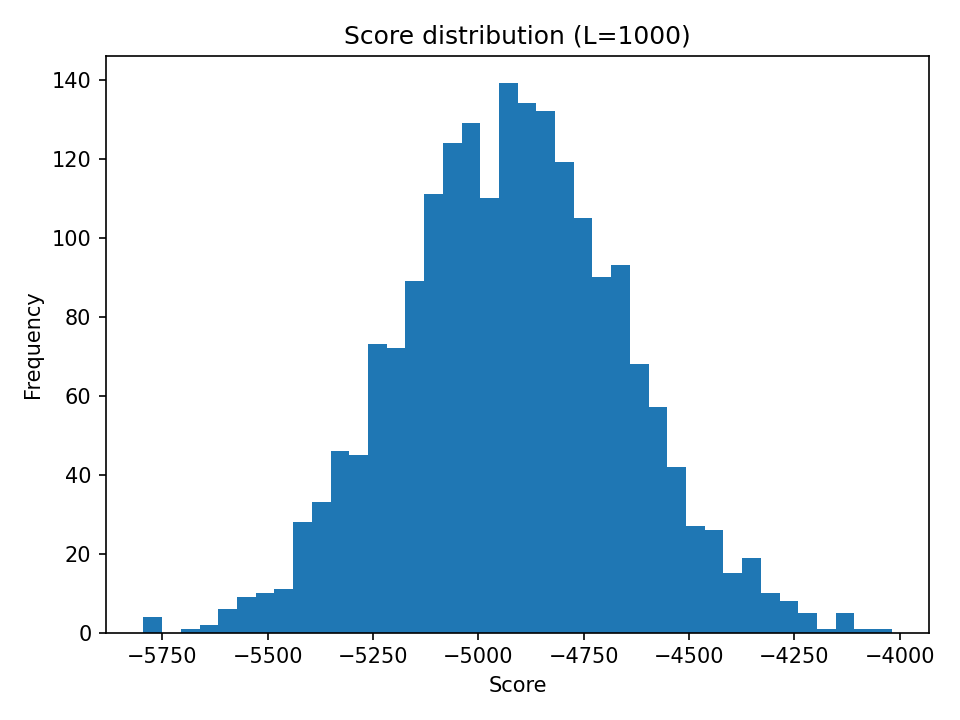}
        \caption{L = 1000}
    \end{subfigure}

    \caption{Distributions of normalized trigram scores for randomly sampled keys at five cipher lengths.}
    \label{fig:score-dists}
\end{figure}

\subsection{Grover-style quantum-inspired sweep}
To compare with idealized quantum search, we computed the expected number of Grover iterations:

\begin{equation}
N_{\text{oracle}} \approx \frac{\pi}{4\sqrt{p_{\text{good}}}}.
\end{equation}

A key is marked whenever $S(\pi) \ge \tau S_{\max}$, and the marked fraction $p_{\text{good}}$ determines the required amplification steps \citep{Boyer1998}. We evaluated all cipher lengths across all thresholds.

\subsection{QUBO annealing framework}
We encoded each permutation $\pi$ as a binary assignment matrix and minimized

\begin{equation}
E(\pi) = -S(\pi) + \lambda C(\pi),
\end{equation}

with $C(\pi)$ enforcing permutation constraints. A simulated quantum annealer with 500 sweeps was used for optimization \citep{dorband2018methodfindinglowerenergy}.

\subsection{Implementation details}
All experiments were implemented in Python using \texttt{NumPy} and \texttt{SciPy}. QUBO optimization used a standard binary representation. Monte Carlo estimates of $p_{\text{good}}$ used 10\,000 samples unless noted otherwise.

\section{Results}
\label{sec:results}

\subsection{Length-dependent concentration of high-scoring keys} \label{subsec1}
Across all corpora, score distributions for randomly sampled permutations become progressively narrower as the cipher length increases (Fig.~\ref{fig:score-dists}). Short ciphers ($200$--$400$ characters) admit a wide range of scores, whereas longer ciphers ($800$--$1000$ characters) produce sharply peaked distributions. This contraction corresponds to a monotonic decrease of $p_{\text{good}}$ across all thresholds.

\begin{figure}[b]
    \centering
    \begin{subfigure}[t]{0.32\textwidth}
        \centering
        \includegraphics[width=\linewidth]{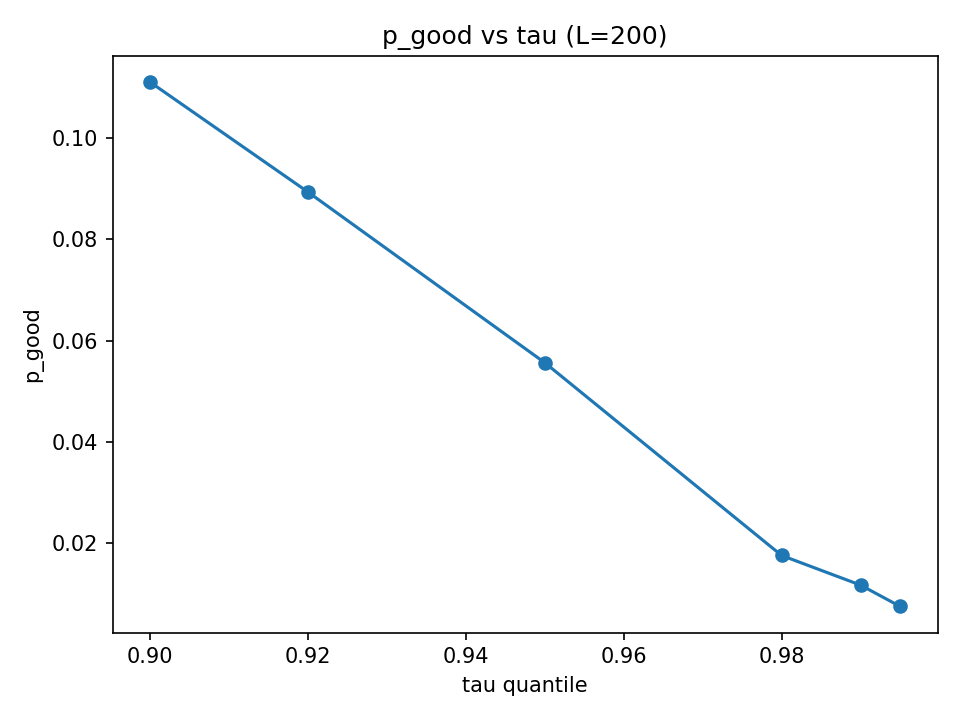}
        \caption{L = 200}
    \end{subfigure}
    \hfill
    \begin{subfigure}[t]{0.32\textwidth}
        \centering
        \includegraphics[width=\linewidth]{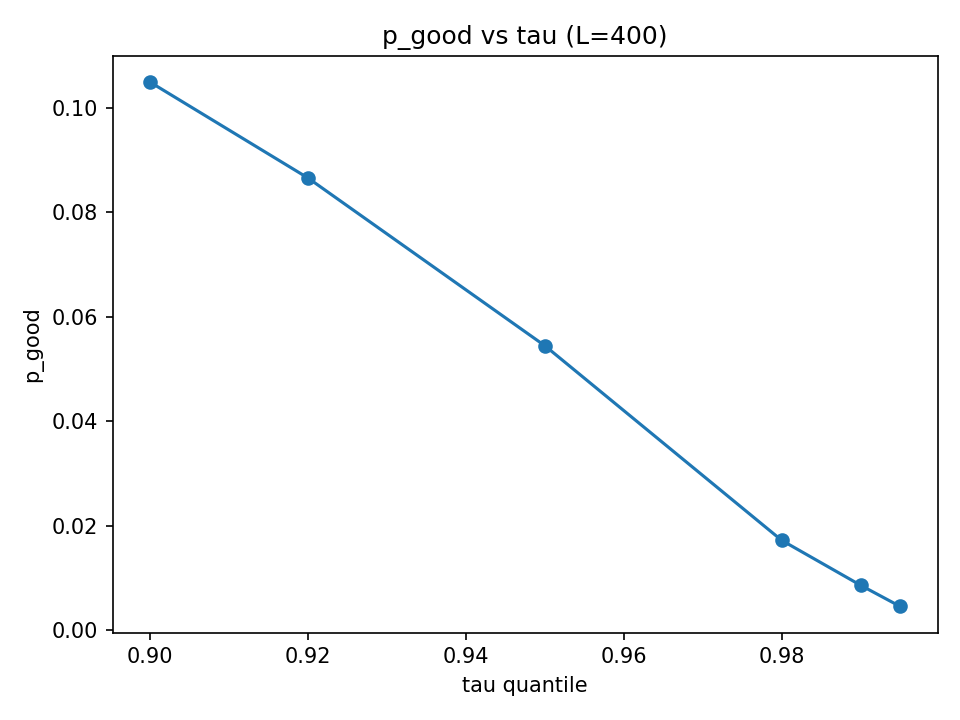}
        \caption{L = 400}
    \end{subfigure}
    \hfill
    \begin{subfigure}[t]{0.32\textwidth}
        \centering
        \includegraphics[width=\linewidth]{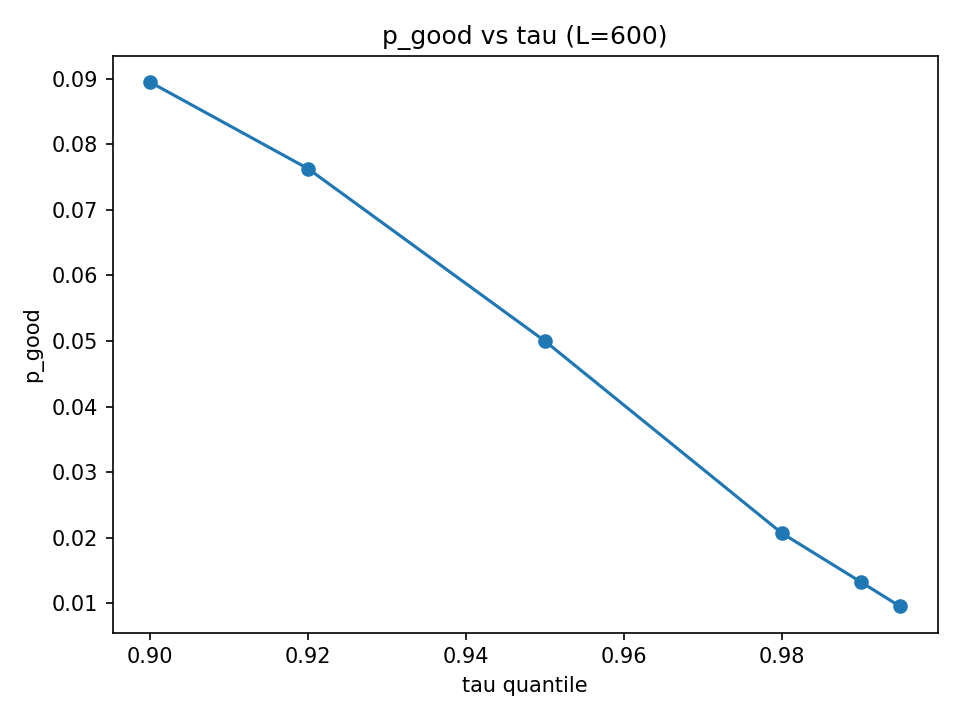}
        \caption{L = 600}
    \end{subfigure}
    
    \vspace{0.7em}

    \begin{subfigure}[t]{0.32\textwidth}
        \centering
        \includegraphics[width=\linewidth]{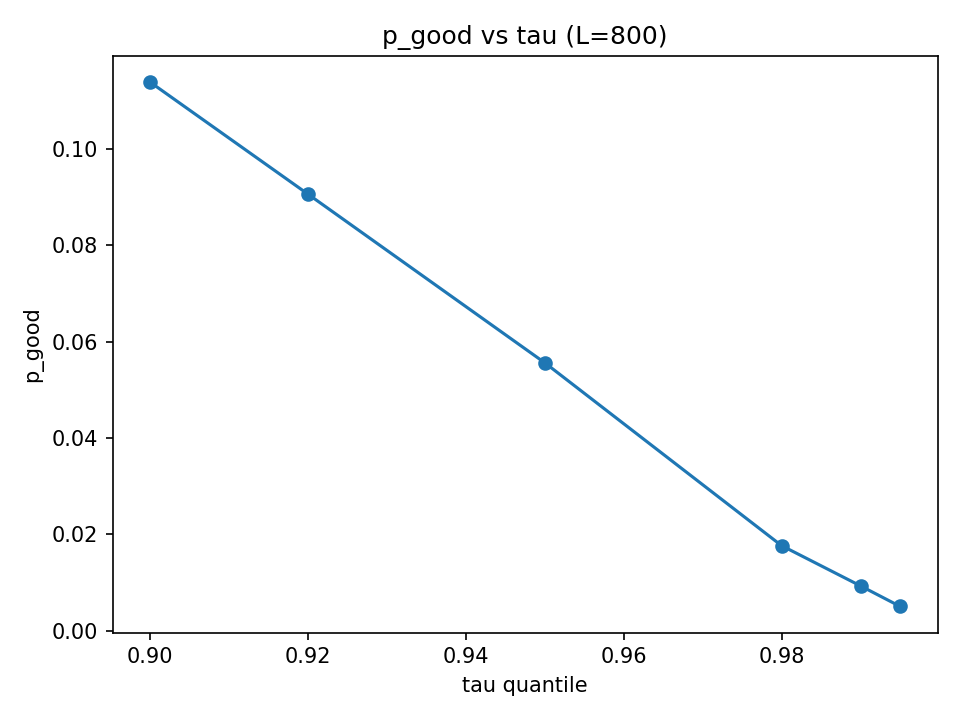}
        \caption{L = 800}
    \end{subfigure}
    \hfill
    \begin{subfigure}[t]{0.32\textwidth}
        \centering
        \includegraphics[width=\linewidth]{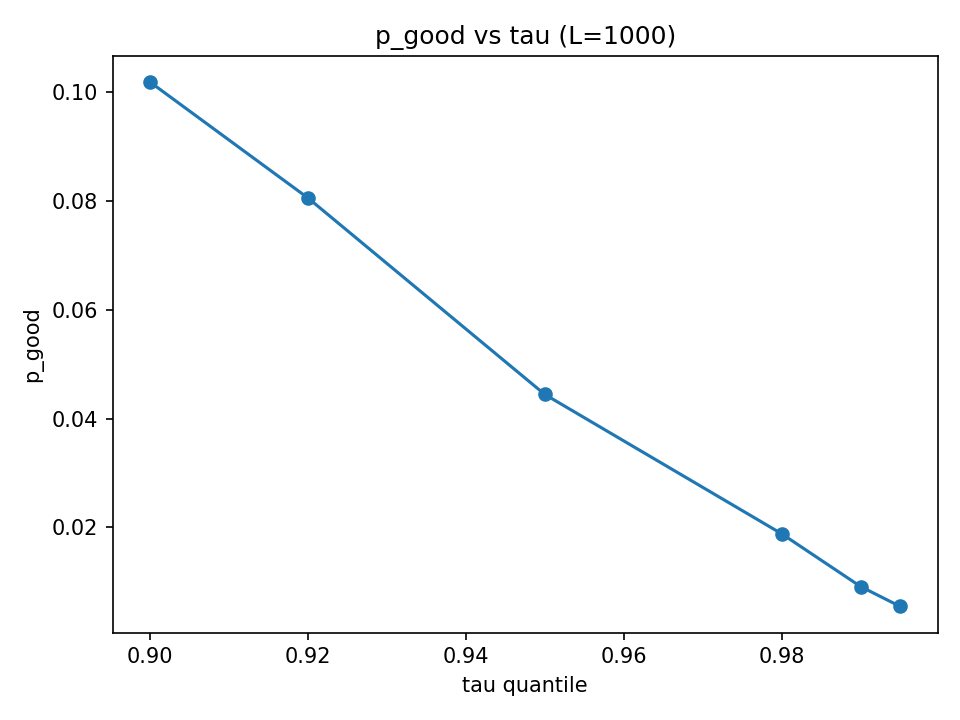}
        \caption{L = 1000}
    \end{subfigure}
    
    \caption{Estimated fraction of good keys $p_{\text{good}}$ as a function of the score threshold $\tau$.}
    \label{fig:pgood-tau-5}
\end{figure}

\subsection{Behavior of $p_{\text{good}}$ across corpora}
Despite substantial stylistic variation, the four corpora yield closely aligned $p_{\text{good}}$ curves across lengths and thresholds (Fig.~\ref{fig:pgood-tau-5}). Political prose (\emph{Machiavelli}, \emph{Guicciardini}) exhibits slightly steeper decay, while \emph{Castiglione} and \emph{Ariosto} show marginally broader curves at intermediate lengths. For all corpora, $p_{\text{good}}$ falls below $10^{-4}$ for $L \ge 600$ when $\tau \ge 0.95$.
\vspace{50pt}
\subsection{Grover-style scaling and oracle iteration estimates} \label{subsec3}
\begin{figure}[t]
    \centering

    \begin{subfigure}[t]{0.26\textwidth}
        \centering
        \includegraphics[width=\linewidth]{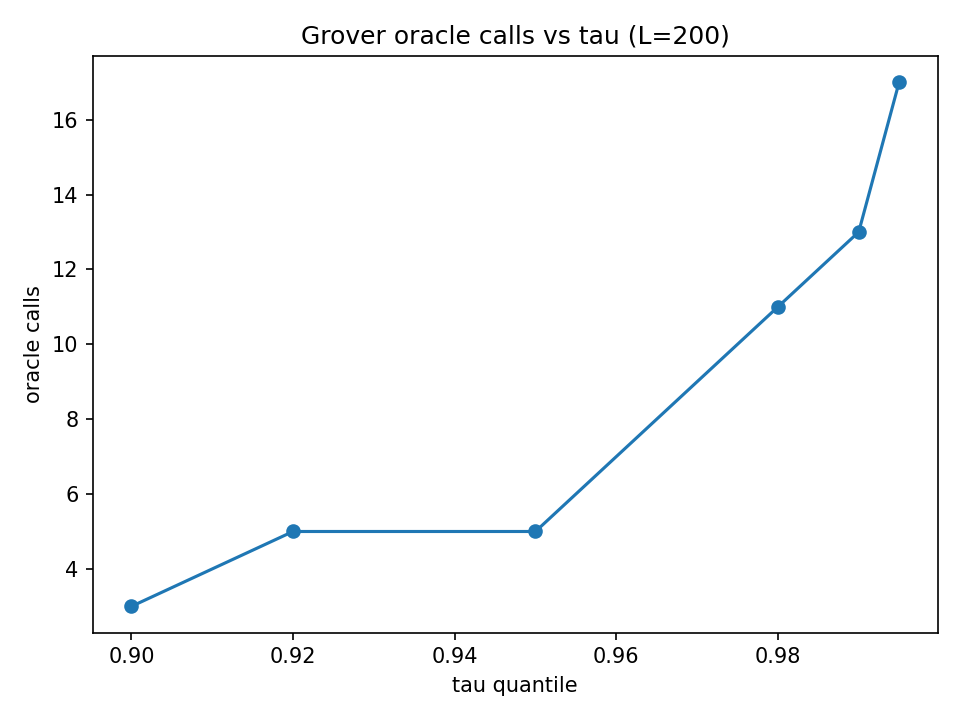}
        \caption{L = 200}
    \end{subfigure}
    \hfill
    \begin{subfigure}[t]{0.26\textwidth}
        \centering
        \includegraphics[width=\linewidth]{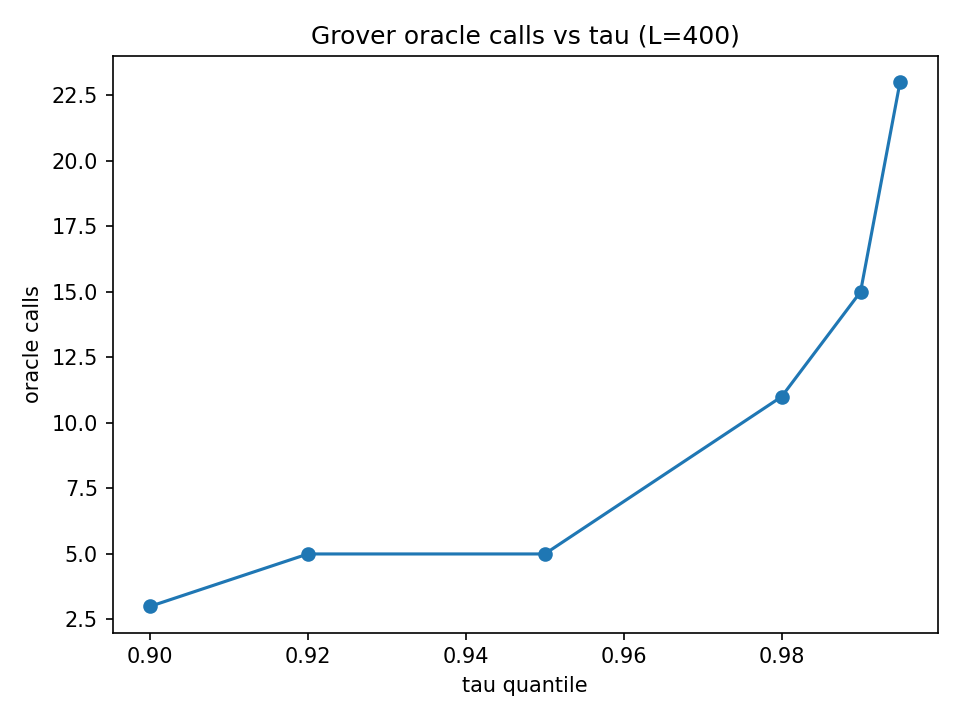}
        \caption{L = 400}
    \end{subfigure}
    \hfill
    \begin{subfigure}[t]{0.26\textwidth}
        \centering
        \includegraphics[width=\linewidth]{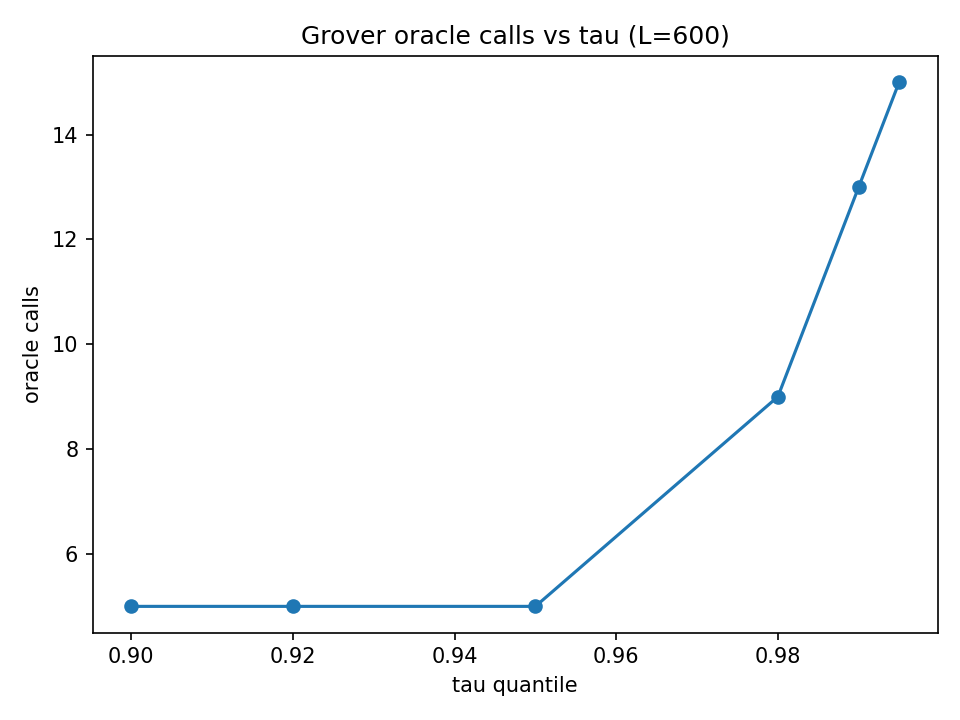}
        \caption{L = 600}
    \end{subfigure}
    
    \vspace{0.7em}

    \begin{subfigure}[t]{0.26\textwidth}
        \centering
        \includegraphics[width=\linewidth]{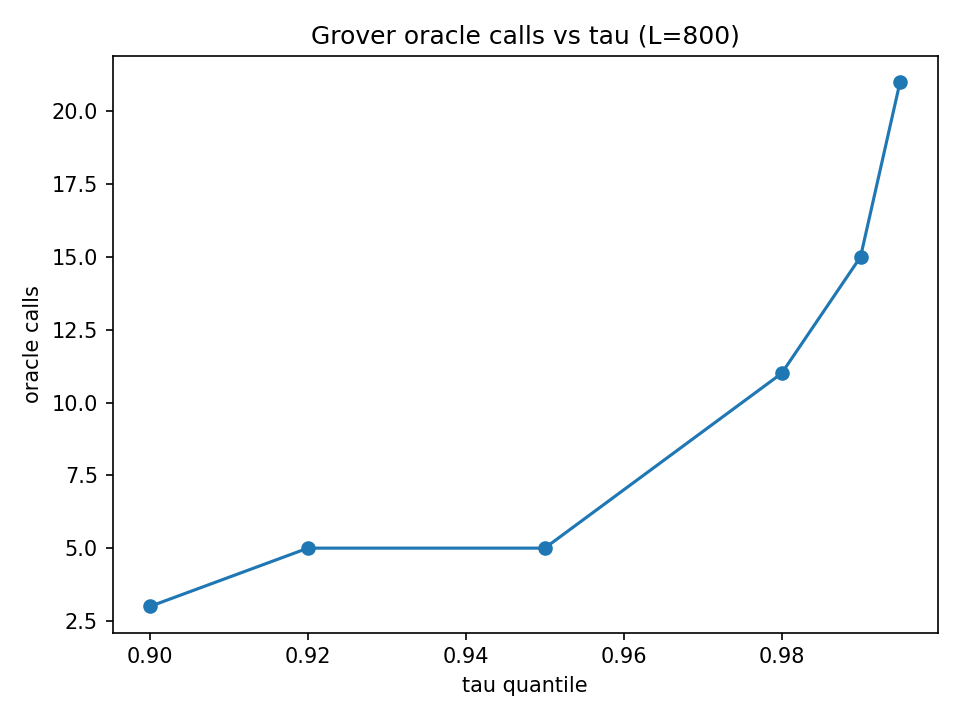}
        \caption{L = 800}
    \end{subfigure}
    \hfill
    \begin{subfigure}[t]{0.26\textwidth}
        \centering
        \includegraphics[width=\linewidth]{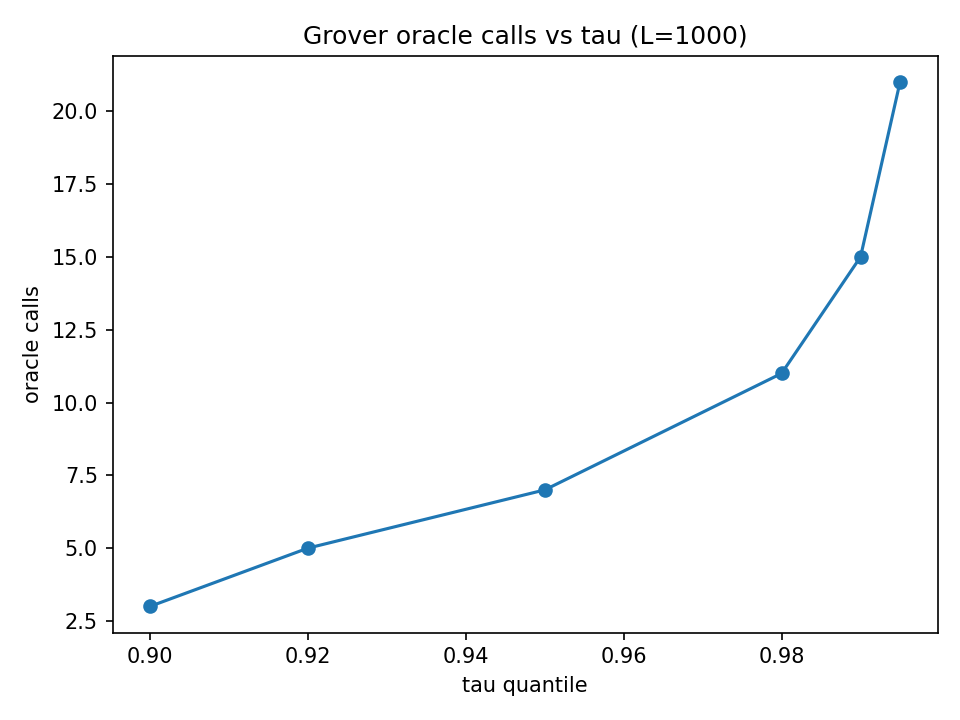}
        \caption{L = 1000}
    \end{subfigure}
    
    \caption{Estimated Grover oracle calls as a function of the threshold $\tau$.}
    \label{fig:oracle-tau-5}
\end{figure}

\begin{figure}[t]
    \centering

    \begin{subfigure}[t]{0.26\textwidth}
        \centering
        \includegraphics[width=\linewidth]{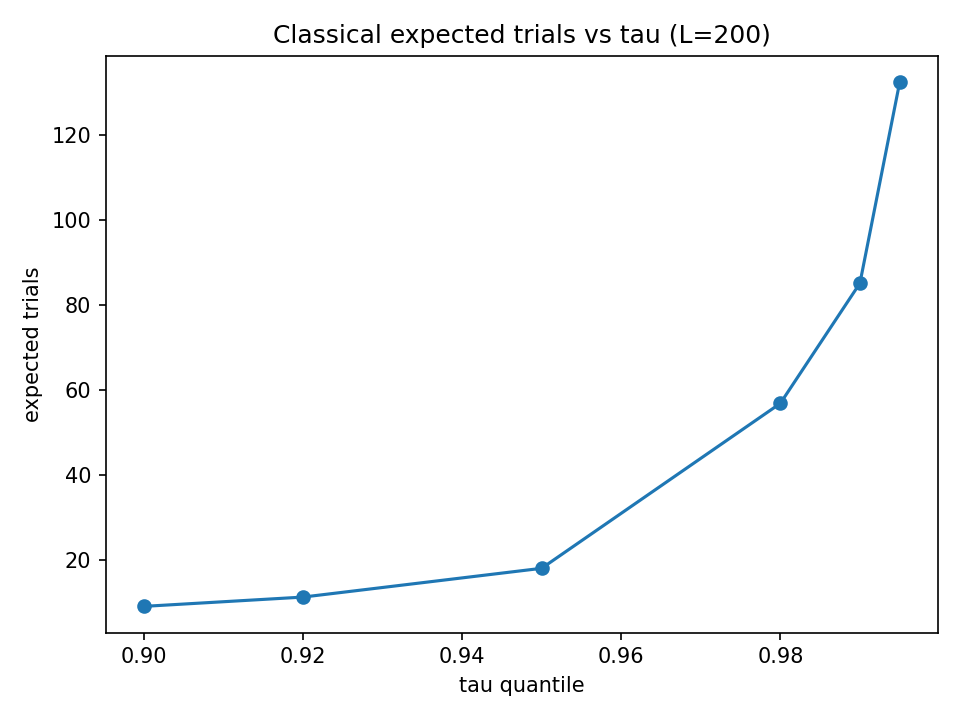}
        \caption{L = 200}
    \end{subfigure}
    \hfill
    \begin{subfigure}[t]{0.26\textwidth}
        \centering
        \includegraphics[width=\linewidth]{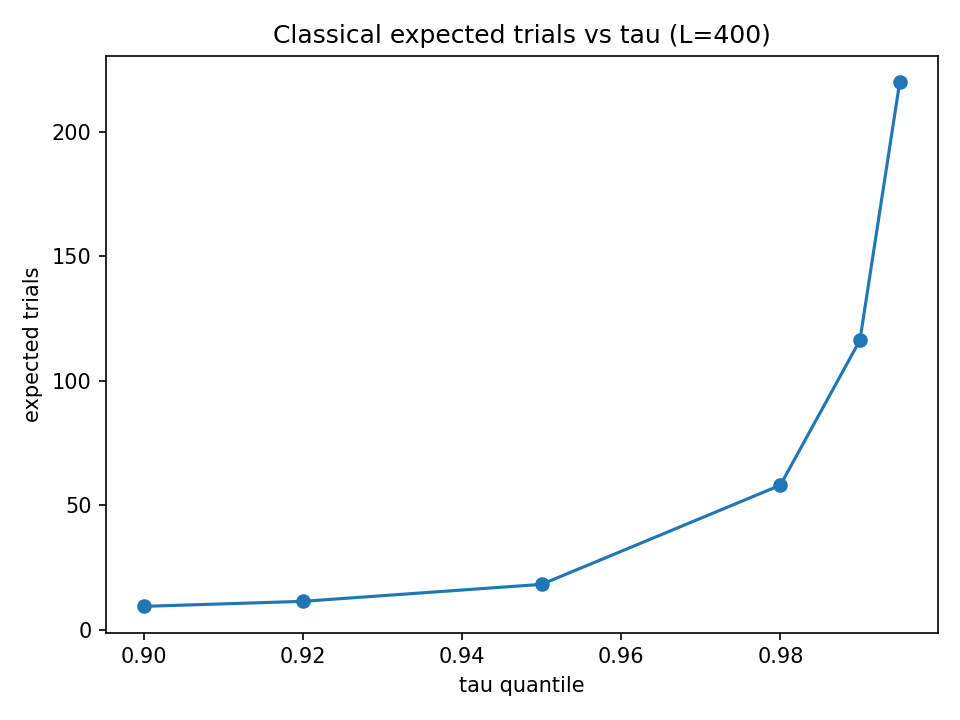}
        \caption{L = 400}
    \end{subfigure}
    \hfill
    \begin{subfigure}[t]{0.26\textwidth}
        \centering
        \includegraphics[width=\linewidth]{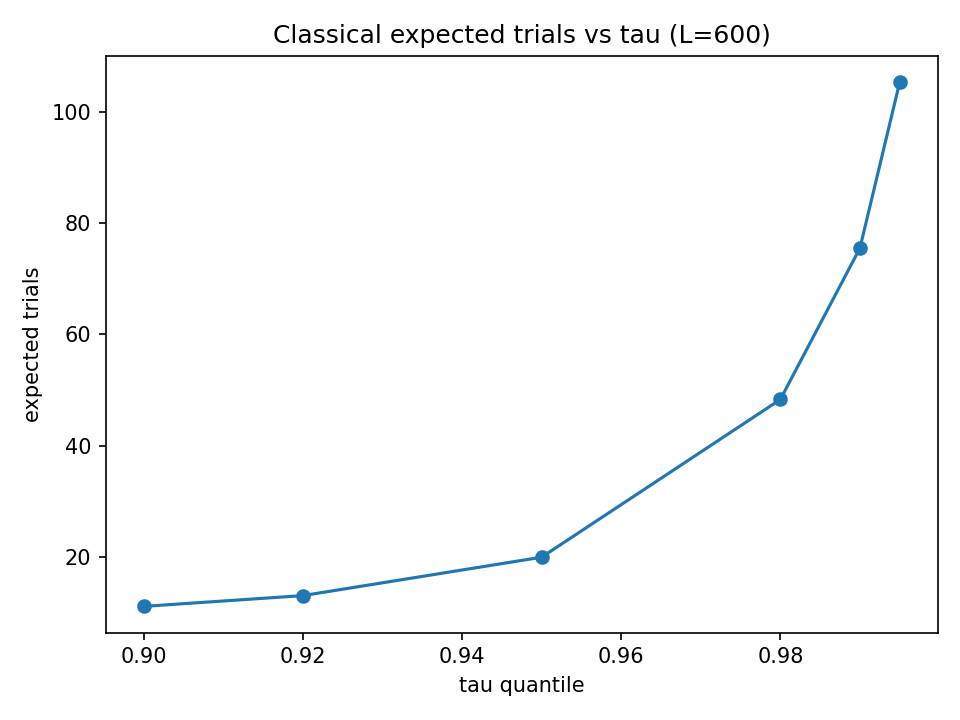}
        \caption{L = 600}
    \end{subfigure}
    
    \vspace{0.7em}

    \begin{subfigure}[t]{0.26\textwidth}
        \centering
        \includegraphics[width=\linewidth]{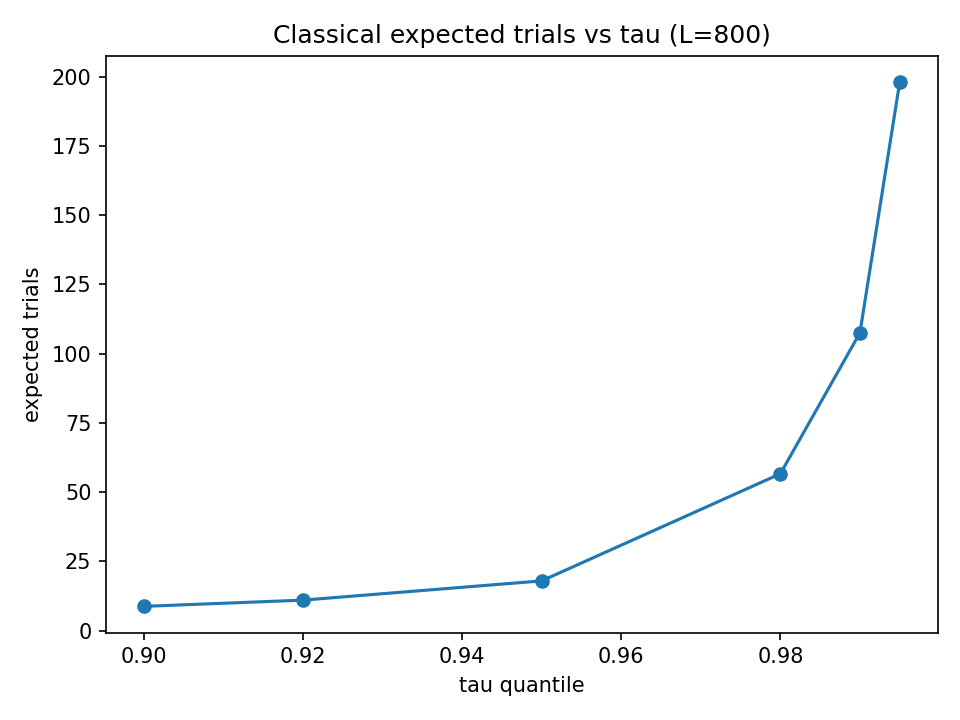}
        \caption{L = 800}
    \end{subfigure}
    \hfill
    \begin{subfigure}[t]{0.26\textwidth}
        \centering
        \includegraphics[width=\linewidth]{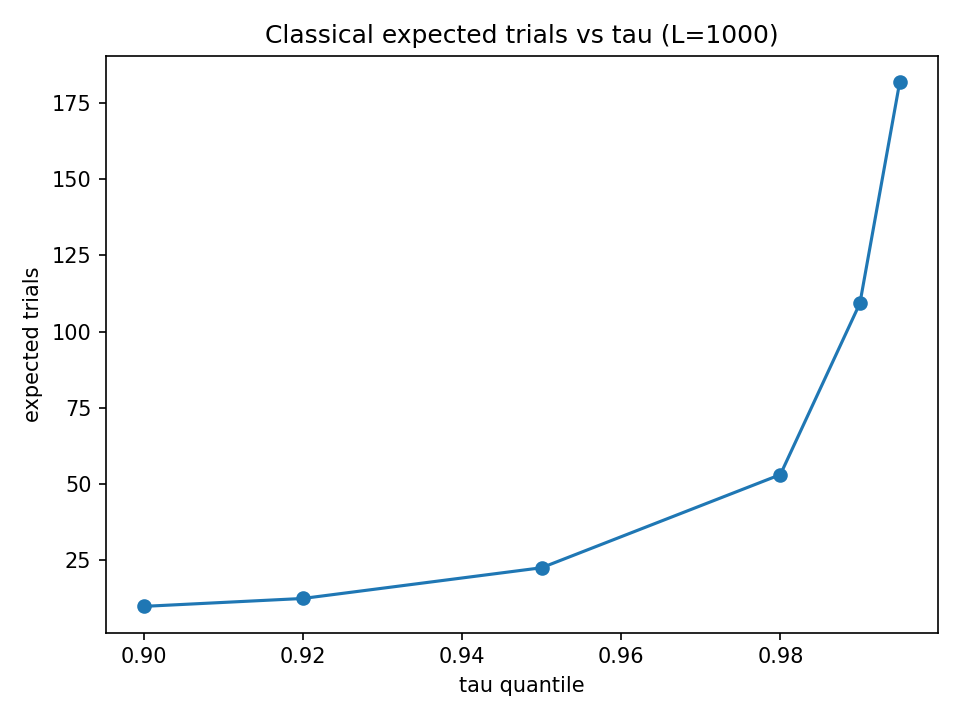}
        \caption{L = 1000}
    \end{subfigure}
    
    \caption{Classical expected number of random trials as a function of the threshold $\tau$.}
    \label{fig:classical-trials-tau-5}
\end{figure}

\begin{figure}[t]
    \centering
    \begin{subfigure}[t]{0.48\textwidth}
        \centering
        \includegraphics[width=\linewidth]{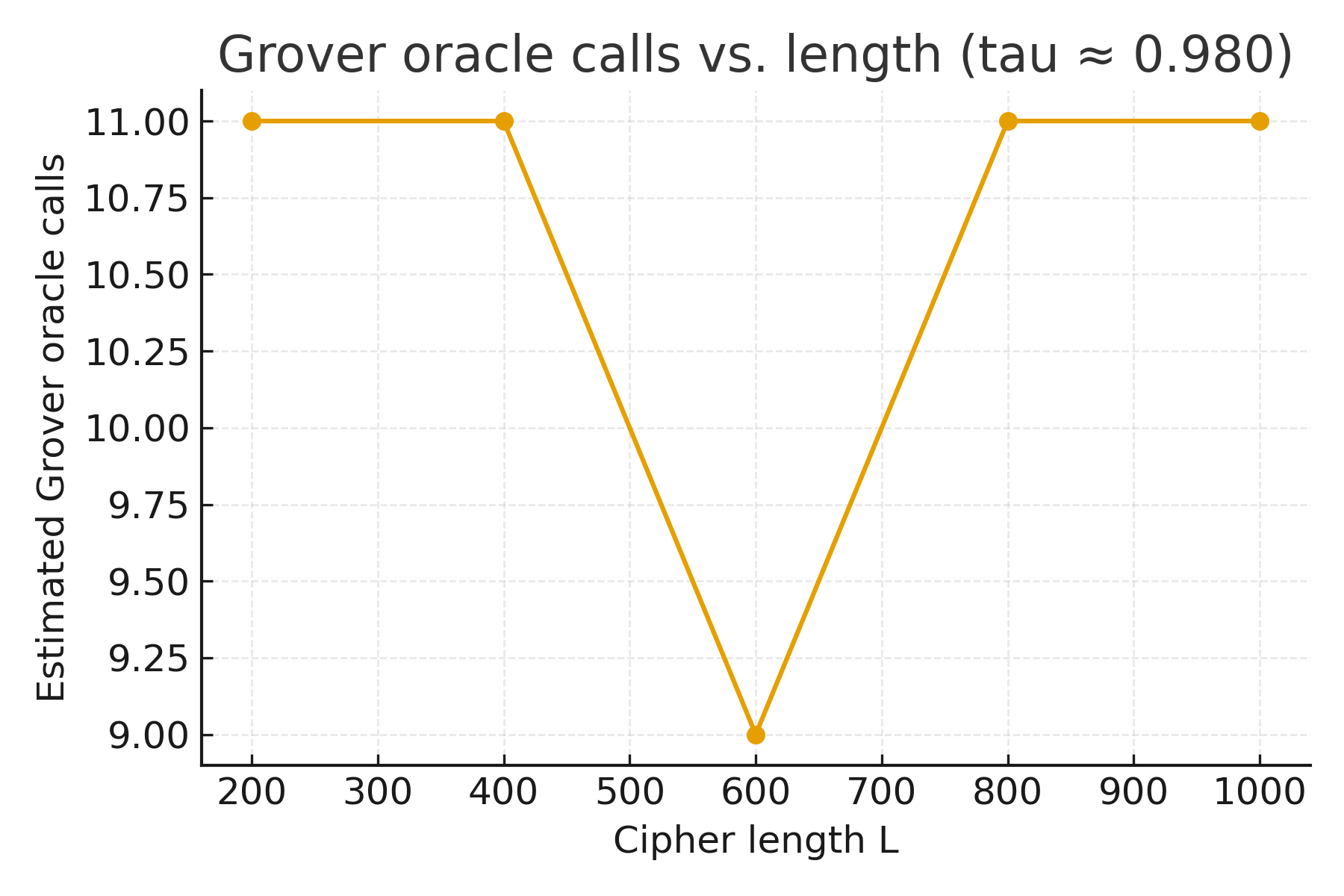}
        \caption{Grover oracle calls vs. length}
    \end{subfigure}
    \hfill
    \begin{subfigure}[t]{0.48\textwidth}
        \centering
        \includegraphics[width=\linewidth]{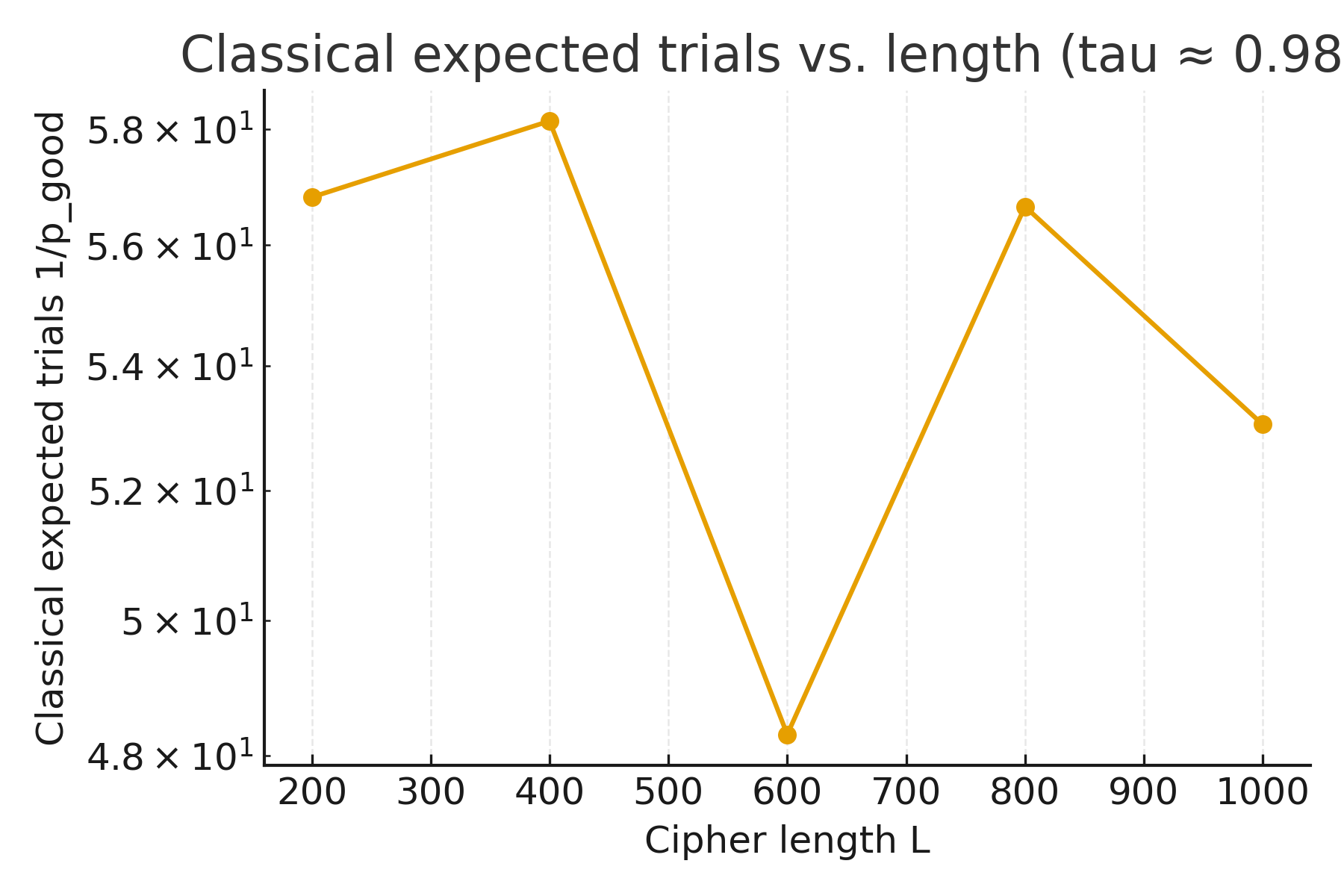}
        \caption{Classical expected trials vs. length}
    \end{subfigure}
    
    \caption{Scaling of search effort with cipher length at a fixed stringent threshold ($\tau \approx 0.98$).}
    \label{fig:length-scaling}
\end{figure}

The expected Grover iteration count scales as $1/\sqrt{p_{\text{good}}}$, and our empirical estimates follow this trend closely (Fig.~\ref{fig:oracle-tau-5}). At stringent thresholds ($\tau \ge 0.98$), required iterations grow rapidly with cipher length, reaching $10^{3}$–$10^{4}$ for $L=600$ and exceeding $10^{5}$ for $L=1000$. Classical expected trials, proportional to $1/p_{\text{good}}$, increase even more sharply (Fig.~\ref{fig:classical-trials-tau-5}). Fig.~\ref{fig:length-scaling} compares both measures at $\tau \approx 0.98$.

\subsection{Comparison with classical heuristic search}
\begin{figure}[t]
    \centering
    \includegraphics[width=0.7\linewidth]{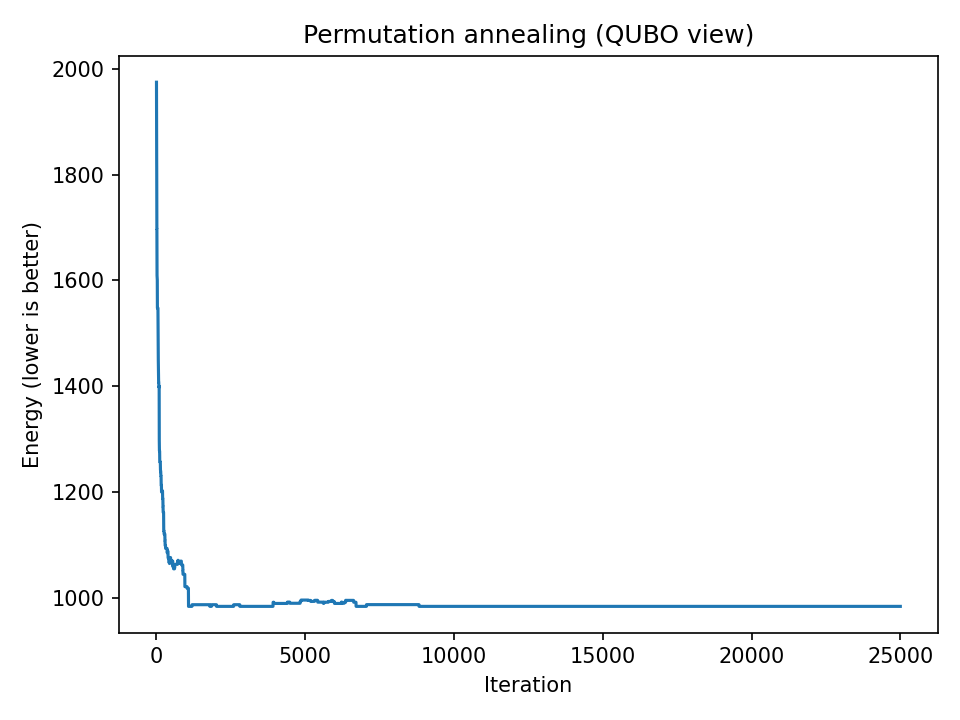}
    \caption{Convergence trajectory of the QUBO-based permutation annealing}
    \label{fig:qubo_energy_curve}
\end{figure}

Before comparing the final scores obtained by hill climbing, simulated annealing, and QUBO-based permutation annealing, we first examine the optimization trajectory produced by the QUBO formulation. Figure~\ref{fig:qubo_energy_curve} shows a representative energy descent curve obtained after simulation. The trajectory exhibits the characteristic pattern of permutation annealing: a rapid initial drop in energy during the early iterations, followed by a long-tail stabilization phase as the search approaches a near-optimal region of the score landscape. This behavior reflects the progressive narrowing of the feasible region discussed in Sections~\ref{subsec1}–~\ref{subsec3}.

\begin{figure}[H]
    \centering
    \includegraphics[width=0.6\textwidth]{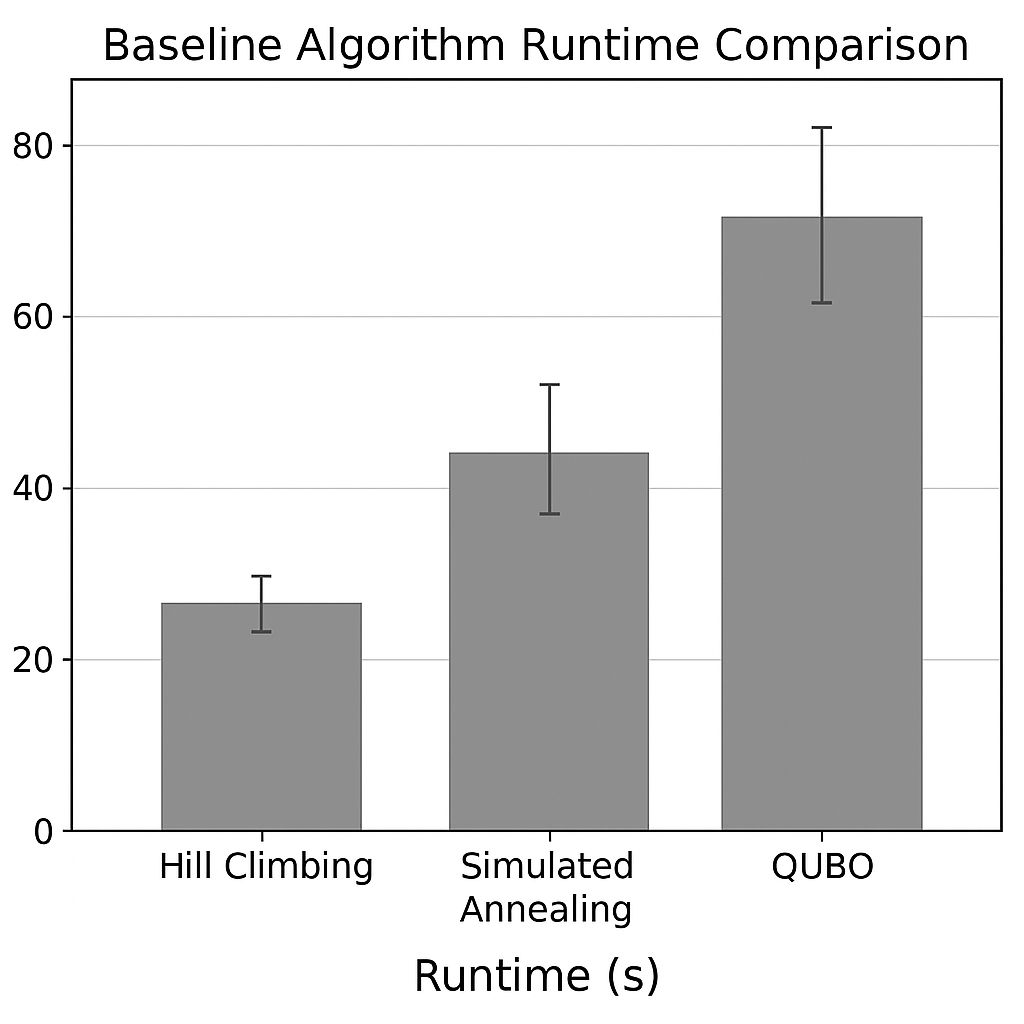}
    \caption{Runtime versus achieved score for classical optimization methods.}
    \label{fig:classical-baselines}
\end{figure}

Building on this, we evaluate the performance of the three classical optimization methods in terms of their runtime and the maximum score they achieve. The results, summarized in Figure~\ref{fig:classical-baselines}, show that all methods experience increasing difficulty as the cipher length grows, mirroring the contraction of the high-scoring region quantified by the decay of $p_{\mathrm{good}}$.

\subsection{Cross-corpus consistency and linguistic interpretation}
The stability of results across genres suggests that, at the character-trigram level, Renaissance Italian forms a relatively homogeneous statistical environment for permutation recovery. The consistent shrinkage of $p_{\text{good}}$ with length reflects a linguistic property rather than an artifact of any one text. This structural regularity underlies both classical and quantum-inspired search behavior.

It is worth noting that the Grover and classical curves in Fig.~\ref{fig:length-scaling} are not strictly monotonic: both exhibit a local dip at $L=600$ before increasing again. This does not contradict the general trend reported in Sections~\ref{subsec1}–~\ref{subsec3}. The value of $p_{\text{good}}$ at each length is an empirical estimate obtained from \textit{Monte-Carlo} sampling over 10\,000 random permutations, and small variations in the upper tail of the score distribution can produce local fluctuations, especially under stringent thresholds such as $\tau \approx 0.98$. In particular, for some corpora the distribution at $L=600$ retains a slightly heavier tail than at $L=400$, resulting in a transient increase in $p_{\text{good}}$ before the expected exponential decay reasserts itself at larger lengths. The overall pattern remains consistent with a sharp contraction of the feasible key space as $L$ increases, even though local deviations from perfect monotonicity arise from finite-sample variability and corpus-specific stylistic effects.

\section{Discussion}
\label{sec:discussion}

This study links corpus-derived linguistic regularities to the effective complexity of a classical cryptanalytic search problem. By estimating the fraction of linguistically plausible keys $p_{\text{good}}$ from historical corpora and relating these estimates to classical, quantum-inspired, and idealized quantum cost models, we show how redundancy in Renaissance Italian shapes the structure of the search space.

\subsection{Linguistic redundancy and search-space contraction}
The empirical $p_{\text{good}}$ curves quantify how linguistic constraints reduce the volume of permutation space compatible with a given ciphertext. Longer texts impose stronger restrictions, leading to sharply peaked score distributions and exponential reductions in $p_{\text{good}}$. Despite differences in genre and style, all four texts exhibit aligned patterns, suggesting that these effects arise from structural features of Renaissance Italian rather than from individual authorial idiosyncrasies.

The small non-monotonic fluctuation observed at $L=600$ reflects finite-sample variability in the empirical estimation of $p_{\text{good}}$ and does not alter the overall exponential decrease of the feasible key region with increasing length.

\subsection{Classical and quantum-inspired search on a unified scale}
Expressing search costs in terms of $p_{\text{good}}$ provides a common basis for comparing classical heuristics, QUBO annealing, and Grover-style search. The observed alignment of Grover oracle estimates with $1/\sqrt{p_{\text{good}}}$ confirms that corpus-driven constraints translate coherently into the quantum search model. Meanwhile, classical heuristics show slowdowns consistent with the shrinking feasible region. These results highlight how search difficulty emerges primarily from linguistic structure.

While character trigrams do not encode long-range syntactic or semantic information, this limitation is acceptable for the task considered here. The cryptanalysis of a monoalphabetic substitution cipher is driven primarily by short-range orthotactic and phonotactic constraints—such as legal consonant clusters, vowel–consonant alternation patterns, and characteristic affixal endings—which trigram models capture effectively. Higher-order $n$-grams would require substantially larger corpora to estimate reliably and would introduce sparsity that distorts $p_{\text{good}}$ estimation, whereas trigrams provide a stable and robust approximation of the local linguistic signal guiding permutation recovery.

\subsection{Methodological scope and limitations}
The present framework employs character trigram models for robustness across medium-sized historical texts. While effective for the cipher task considered here, such models do not capture longer-range syntactic or semantic dependencies, which may further restrict feasible decryption candidates. The cryptanalytic setting is idealized in that we consider clean monoalphabetic substitution and a perfect marking oracle for Grover-style analysis. More complex historical ciphers may yield different $p_{\text{good}}$ profiles.

\subsection{Future directions}
Future work may extend this framework to other languages, historical periods, or stronger linguistic models. Comparing $p_{\text{good}}$ across typologically or diachronically diverse corpora could reveal how linguistic structure interacts with combinatorial search difficulty. From a computational perspective, grounding quantum and classical search analyses in corpus-derived distributions may inform practical benchmarks for language-related optimization tasks.

\section{Conclusion} \label{sec:conclusions}

By integrating corpus-based language modeling with classical, quantum-inspired, and idealized quantum search analyses, this study provides a quantitative account of how linguistic regularities shape permutation recovery in monoalphabetic substitution ciphers. The empirical estimates of $p_{\text{good}}$ reveal that Renaissance Italian imposes strong and systematic constraints on the search space, yielding consistent patterns across authors and genres. These constraints manifest directly in both classical heuristic performance and Grover-style oracle estimates.

Beyond the specific cryptanalytic task, the framework demonstrates how quantitative linguistics can inform computational complexity by connecting redundancy in historical texts to measurable search dynamics. This opens avenues for cross-linguistic comparisons, richer linguistic models, and further integration of corpus-driven methods with analyses of classical and quantum search processes.

\section*{Data Availability Statement}
The data that support the findings of this study are openly available in \textit{Quantitative-and-Quantum-Inspired-Analysis} at \href{https://github.com/alessiobb3b/Quantitative-and-Quantum-Inspired-Analysis}{GitHub}.

\section*{Declarations}
None of the authors have a conflict of interest to disclose.

\bibliography{sn-bibliography}

\end{document}